# QUANTISING GENERAL RELATIVITY USING QED THEORY

*AN OVERVIEW AND EXTENSION*

S.B.M. Bell[†]

Interdisciplinary Quantum Group, Department of Computer Science

The University of Liverpool, Chadwick Building, Peach Street

Liverpool, L69 7ZF, United Kingdom

## Abstract

We summarise and discuss some of our previous results,[7-10] which show that Bohr's theory of the one-electron atom[13, 14] may be derived from the theory underpinning Quantum ElectroDynamics (QED)[18] or vice versa, and that General Relativity[19] may also be derived from QED theory in the classical limit, if we use Newtonian mechanics[21, 24] in the right frame and self-similar tesseral hierarchies.[4-6, 17] We circumvent Newton's arguments[24] against Descartes' vortex theory[16] to show that the inverse square law for a force[24] combined with the equation of circular motion[24] and Bohr's quantisation of angular momentum[13, 14] may be derived from the vortex theory[16, 24] and Special Relativity.[19] We remark on the electro-weak interaction, the number of dimensions needed, and their connection with tesseral hierarchies.[4-6, 8, 10, 17]

---

[†] Mail@sarahbell.org.uk
.

## 1. Notation

Matrices with more than one row or column and quaternions are given boldface type. * signifies complex conjugation. $^T$ signifies transposition. $^\dagger$ signifies Hermitian conjugation. $^\ddagger$ signifies quaternion conjugation.[1, 21] A lowercase Greek subscript stands for 0, 1, 2, or 3 and indicates the spacetime axes. $i = \sqrt{-1}$. $\mathbf{i}_0 = 1$. $\mathbf{i}_1 = \mathbf{i}$, $\mathbf{i}_2 = \mathbf{j}$, $\mathbf{i}_3 = \mathbf{k}$ stand for the quaternion matrices, where $\mathbf{i}_r^2 = -1$ and $\mathbf{i}_1\mathbf{i}_2 = \mathbf{i}_3$, $\mathbf{i}_2\mathbf{i}_1 = -\mathbf{i}_3$ with cyclic variations.[1, 21] We have $\mathbf{i}_1* = -\mathbf{i}_1$, $\mathbf{i}_2* = \mathbf{i}_2$, $\mathbf{i}_3* = -\mathbf{i}_3$, $\mathbf{i}_1^T = \mathbf{i}_1$, $\mathbf{i}_2^T = -\mathbf{i}_2$, $\mathbf{i}_3^T = \mathbf{i}_3$, $\mathbf{i}_r^\dagger = -\mathbf{i}_r$ and $\mathbf{i}_r^\ddagger = -\mathbf{i}_r$. We note that $\mathbf{i}_0* = \mathbf{i}_0^T = \mathbf{i}_0^\dagger = \mathbf{i}_0^\ddagger = 1$ and that $i^\ddagger = i \neq i^\dagger = -i$. We define the underline symbol and the vertical bar as indicating *a reflector* or *a rotator* matrix,[7]

$$\underline{\mathbf{U}} = \underline{\mathbf{U}}(\mathbf{Q}, \mathbf{U}) = \begin{pmatrix} 0 & \mathbf{Q} \\ \mathbf{U} & 0 \end{pmatrix}, \quad \mathbf{U}| = \mathbf{U}|(\mathbf{Q}, \mathbf{U}) = \begin{pmatrix} \mathbf{Q} & 0 \\ 0 & \mathbf{U} \end{pmatrix}$$

We set the speed of light, c, and the gravitational constant, *G*, to one.

## 2. The four-vector spinor

We have shown that the particle version of QED theory applied to electromagnetism, comprising the Dirac equation[18] and Maxwell's equations[12] for the photon, can be reproduced with the spinor behaving like a four-vector.[7] We shall call Maxwell's equations collectively, in the form of an equation for the potential, the radiation equation.

The Dirac equation in the form required for demonstrating this behaviour is[7]

$$\left(\left(h/2\pi\right)\underline{\mathbf{D}}\left(\mathbf{D}, \mathbf{D}^{\ddagger}\right) - ie\,\underline{\mathbf{A}}^{\sim}\left(\mathbf{A}^{\sim}, \mathbf{A}^{\sim\ddagger}\right)\right)\underline{\Phi}\left(\phi_1, \phi_2\right) = \qquad (2.A)$$
$$\underline{\Phi}\left(\phi_1, \phi_2\right)\underline{\mathbf{M}}\left(\mathbf{M}, -\mathbf{M}^{\ddagger}\right)$$

where

$$\mathbf{A}^{\sim} = -i\mathbf{i}_0 A_0 + \mathbf{i}_1 A_1 + \mathbf{i}_2 A_2 + \mathbf{i}_3 A_3,$$
$$\mathbf{D} = -i\,\partial/\partial x_0 + \mathbf{i}_1\,\partial/\partial x_1 + \mathbf{i}_2\,\partial/\partial x_2 + \mathbf{i}_3\,\partial/\partial x_3,$$
$$\phi_1 = \left(\psi_1 + \psi_2\right)\left(1 + i\mathbf{i}_3\right),$$
$$\phi_2 = i\left(\psi_1 - \psi_2\right)\left(1 + i\mathbf{i}_3\right),$$
$$\mathbf{M} = m_e/i$$

$\left(A_0, A_1, A_2, A_3\right)$ is the potential, $\psi_1$ and $\psi_2$ are quaternion representations of the usual bispinors, $\psi_1$ and $\psi_2$.[18, 21] $m_e$ and $e$ are the mass and charge of the particle in any one chosen frame. If $\underline{\Phi}$ transforms in the usual way, so that the phase, $\phi$, traverses half the angle rotated and the analogue under Lorentz transformation, $\mathbf{M}$ must remain a constant scalar and the Dirac equation, (2.A), is simply another form of the usual version. However, if we take $\mathbf{M}$ as the energy-momentum four-vector, $\underline{\Phi}$ must behave like a four-vector as well for Special Relativity to be observed, with $\phi$ traversing the whole angle.

Transforming to a dashed frame[7] and using $\underline{\Phi}$ as an example,

$$\underline{\Phi}' = \mathbf{R}|\,\underline{\Phi}\,\mathbf{R}|^{\ddagger n}$$

where $n = 0$ for half-angular behaviour and $n = 1$ for four-vector behaviour, and,

$$\mathbf{R}| = \mathbf{R}_S|\left(\mathbf{R}, \mathbf{R}\right), \quad \mathbf{R}| = \mathbf{R}_T|\left(\mathbf{R}, \mathbf{R}^{\ddagger}\right)$$

for a spatial rotation and a Lorentz transformation, respectively, with

$$\mathbf{R} = \cos\frac{\phi}{2} + \mathbf{i}_1 \sin\frac{\phi_1}{2} + \mathbf{i}_2 \sin\frac{\phi_2}{2} + \mathbf{i}_3 \sin\frac{\phi_3}{2},$$

$$\sin^2\frac{\phi_1}{2} + \sin^2\frac{\phi_2}{2} + \sin^2\frac{\phi_3}{2} = \sin^2\frac{\phi}{2}$$

where $\phi$ and $\phi_r$ are real in the former case and imaginary in the latter. They are closely related to the Cayley-Klein parameters.[1, 7, 21]

The Dirac equation, (2.A), has a conserved current, $\mathbf{J}_\mu^\S$,[7]

$$i\mathbf{J}_0^\S = t : \{\mathrm{Trace}(\underline{\mathbf{K}}_0 \underline{\boldsymbol{\Phi}}_S \underline{\mathbf{I}}_0 \underline{\boldsymbol{\Phi}})\}, \quad \mathbf{J}_r^\S = t : \{\mathrm{Trace}(\underline{\mathbf{K}}_r \underline{\boldsymbol{\Phi}}_S \underline{\mathbf{I}}_r \underline{\boldsymbol{\Phi}})\}, \quad (2.\mathrm{B})$$

$$\underline{\boldsymbol{\Phi}}_S = \underline{\boldsymbol{\Phi}}_S\left(\boldsymbol{\phi}_1^\dagger, \boldsymbol{\phi}_2^\dagger\right) \quad \underline{\mathbf{I}}_\mu = \underline{\mathbf{I}}_\mu\left(\mathbf{i}_\mu, \mathbf{i}_\mu^\ddagger\right) \quad \underline{\mathbf{K}}_\mu = \underline{\mathbf{K}}_\mu\left(\mathbf{k}_\mu, \mathbf{k}_\mu^\ddagger\right)$$

where $t :$ stands for extracting the *temporal* or scalar part of the expression, and $\mathbf{k}_\mu = -ie/4$ in the chosen frame. The radiation equation using $\mathbf{J}_\mu^\S$ allows the same half- or whole-angular behaviour of $\phi$.

## 3.    Bohr's equations and electromagnetic particles

This leads to the question: "Can QED theory be applied in the classical case?" We tried Bohr's semi-classical theory of the one-electron atom[13, 14] in part 1 of Bell et al.,[8] which we will call part 1. Bohr defined his model using two equations for the charged particle. These were,

(i)    the particle obeys Newton's equation of circular motion [24] in the frame of the source of attraction,

(ii)    the angular momentum of the particle, again in the frame of the source, must be an integer multiple of $h/2\pi$.

We used a version of Bohr's theory [14] in which the particle's mass was allowed to vary in conformity with Special Relativity. Bohr's equations are then

$$\frac{Ze^2}{R} = \frac{m_e v_e^2}{\sqrt{1 - v_e^2}}, \quad \frac{m_e v_e R}{\sqrt{1 - v_e^2}} = \frac{n_\theta h}{2\pi}$$

(3.A)

in the frame of the source, where $Ze$ is the charge on the source, $R$ is the distance between the particle and the source, *the Bohr radius*, $v_e$ is the velocity of the particle and $n_\theta$ is an integer.

We introduced *the Circular transformation*,[8] which turned linear motion into circular motion and an inverse-distance potential into a constant one. This cylindrical symmetry was reached by pinching lines parallel to the *x*-axis into radii meeting at the origin, while lines parallel to the *y*-axis became the corresponding circles. After performing the Circular transformation, the Dirac equation, (2.A), was solved to give

$$\phi_1 = \exp\{i(-\nu x_0 + \mu s)\},$$
$$\phi_2 = \{-(h/2\pi)\nu + i\mathbf{i}_2(h/2\pi)\mu + eA_0\} \times$$
$$\mathbf{M}^{-1} \exp\{i(-\nu x_0 + \mu s)\},$$
$$-m_e^2 = -\{-(h/2\pi)\nu + eA_0\}^2 + (h/2\pi)^2\mu^2$$

where $\nu$ is the frequency and $\mu$ the wave number of the particle, $s$ labels the arc and $A_0$ is the potential in the frame of the source. Using a Lorentz transformation from the frame of the particle in the space formed by the Circular transformation, we showed that the motion satisfied Bohr's first equation, (3.A). Bohr's second equation followed from the requirement that $\underline{\Phi}$ be single-valued for spatial co-ordinates. This only held in general if $\underline{\Phi}$ behaved like a four-vector.

We proved the inverse in part 2 of Bell et al.,[8] which we will call part 2. Using Bohr's equations, we found a set of solutions of the Dirac and radiation equations describing miniature Bohr atoms at each point in spacetime. Each atom was composed of a small part of the wave function of the particle and the charge density of the source. The relation between the variables at a point, was

$$\frac{\rho^2}{ge^2} - A_0^3 \rho - m_e^2 g A_0^4 = 0, \quad g = \frac{3\pi}{n_\theta^2 h^2}$$

in the local rest frame of the source, where $\rho$ is the charge density of the source. We ordered and scaled the atoms into a lattice of spacing $a$, imposing a common frame and allowed the size of the lattice cells to tend to a point. One possible outcome for the order of magnitude was

$$|\mathbf{A}| \sim a^2, \quad e \sim 1, \quad m_e \sim \frac{1}{a^{3+p}}, \quad p > 1$$

where $\mathbf{A}$ is the potential. The complete solution for particular boundary conditions could be found by solving the Dirac and radiation equations.

## 4.    Bohr's equations and tachyons

Bohr's equations, (3.A), only give the principal energy levels. Sommerfeld supplied the missing fine structure by admitting elliptical orbits.[25] This model gives the same energy levels as Dirac obtained by solving the Dirac equation,[18] although each gives different labels for them. A circular projection may be obtained from an ellipse, and we chose to use circles.[9] We considered a circular orbit split into two circular components occupying orthogonal planes in spacetime.[9, 10] The circumference of each component could be thought of as the orbit of a Bohr atom in its own right. The first atom was formed from the original particle and a source in the previous way. The circular temporal co-ordinate formed the orbit of a second, notional, atom, whose matching temporal co-ordinate was the orbit of the original atom. Both the Dirac and radiation equations are invariant when transformed to a frame in which a particle becomes a tachyon, as in the second atom, and this continues to be true when $\underline{\Phi}$ transform like a four-vector. We added the contributions made to the energy by the original atom and the copy in the frame of the source,[9]

$$ \nu + \nu_t = m_e \left( \sqrt{1 - \frac{4\pi^2 e^4}{n_\theta^2 h^2} + \frac{n_r}{n_\theta}} \right) \qquad (4.\text{A}) $$

where $n_r$ is the equivalent of $n_\theta$ for the copy. We show in section 8 that we may derive the inverse square law for the force from Bohr's second equation alone, which we obtained for the copy.[10] The original and the copy could be combined into a structure showing spatial spherical symmetry, with a new, third, temporal co-ordinate.[9] With this temporal co-ordinate, equation (4.A) described the spectrum we see,

$$ \nu + \nu_t \rightarrow m_e \left\{ 1 + \frac{4\pi^2 e^4 / h^2}{\left( \sqrt{n_\theta^2 - 4\pi^2 e^4 / h^2} + n_r \right)^2} \right\}^{-\frac{1}{2}} $$

We extended our previous derivation of QED theory in part 2 to include the copy, showing that this too leads to QED theory.[9]

## 5. Bohr's equations from a metric

We examined whether General Relativity could be expressed in this language.[10] We described the circular motion of a particle in a curved two-dimensional space, later adjusted to four dimensions using spherical symmetry. The metric became

$$\mathrm{d}\tau^2 = \frac{2m_s}{r}\mathrm{d}s^2 - \frac{r}{2m_s}\mathrm{d}r^2 - r^2\left(\mathrm{d}\vartheta^2 + \sin^2\vartheta\,\mathrm{d}\varphi^2\right) \qquad (5.A)$$

where $s$ is the temporal co-ordinate, $r$ is the distance of the particle from the source and $m_s$ is the mass of the source. We showed that motion in a flat space, the *moving space*, could be substituted for curvature with the assignments

$$\hat{\theta} = \theta\sqrt{2}, \quad v_g = \sqrt{\frac{m_s}{r}}, \quad \frac{\mathrm{d}^2 r}{\mathrm{d}\tau^2} = \frac{m_s}{r^2} \qquad (5.B)$$

We define the variables in the moving space. $\tau$ is the temporal co-ordinate rather than the interval, $r$ is the distance of the particle from the source, $m_s$ is the mass of the source, $v_g$ is the velocity of the particle and $\hat{\theta}$ is the angle swept out, all in the frame of the particle. The last equation represents Newton's law of gravity. [24] $\theta$ is the angle swept out in the curved space.

We showed that a wave could accompany the particle[10] if $v_g$ was rational. Then, from equations (5.B),

$$\sqrt{\frac{m_s}{r}} = \frac{m}{n_\theta}$$

where $m$ is an integer. This limited the values of $r$ to a discrete series, $r = R$. As $m_s$ or $R$ grew larger, the behaviour became approximately continuous. All this could be summarised by Bohr's equations in the frame of the source,

$$\frac{m_p \tilde{m}_s}{R} = \frac{m_p}{\sqrt{1-v_g^2}} v_g^2, \quad \frac{m_p}{\sqrt{1-v_g^2}} v_g R = \frac{n_\theta h_g}{2\pi}, \quad (5.C)$$

$$\tilde{m}_s = \frac{m_s}{\sqrt{1-v_g^2}}, \quad \frac{h_g}{2\pi} = \frac{m_p \tilde{m}_s}{m}$$

where $h_g$ is the gravitational version of Planck's constant, which we discuss in sections 6 and 8.

## 6.    Hierarchical Bohr's equations and gravity

We showed that the interaction[22] inside the thin gravitational shell predicted by General Relativity[19] could also be described by Bohr's equations in a flat moving space and that the angle swept out was $\theta$.[10] We ascribed the interaction in the last section to the outside of the shell where it could be seen as a tachyonic transformation of the internal one. Since both can be described using a moving space, the outside could also be seen as the inside of a another shell with the same centre. Iterating, we may imagine the second shell surrounded by a third and so on and the inside containing another shell, which contained another shell and so on, like an infinite series of Russian dolls. Our results could be tabulated so that part of the infinite series reads:

**Hierarchy showing the many guises of a single shell**

| Space no. | Mass | Name | Speed | Q. nos. | $h_g/2\pi$ |
|---|---|---|---|---|---|
| $-1$ | $m_s/4$ <br><br> $\dfrac{m_s/2}{\sqrt{1-v_g^2/4}}$ | protoshell, external <br> multishell, internal | $v_g/2$ | $n$ <br><br> $\dfrac{m}{2}$ | $\dfrac{m_s^2/4}{m\left(1-v_g^2/4\right)}$ |
| $0$ | $m_s/2$ <br><br> $\dfrac{m_s}{\sqrt{1-v_g^2/2}}$ | multishell, external <br> copyshell, internal | $v_g/\sqrt{2}$ | $n$ <br><br> $\dfrac{m}{\sqrt{2}}$ | $\dfrac{m_s^2/\sqrt{2}}{m\left(1-v_g^2/2\right)}$ |
| $1$ | $m_s$ <br><br> $\dfrac{2m_s}{\sqrt{1-v_g^2}}$ | copyshell, external <br> unishell, internal | $v_g$ | $n$ <br><br> $m$ | $\dfrac{2m_s^2}{m\left(1-v_g^2\right)}$ |
| $2$ | $2m_s$ <br><br> $\dfrac{4m_s}{\sqrt{1-2v_g^2}}$ | unishell, external <br> combishell, internal | $v_g\sqrt{2}$ | $n$ <br><br> $m\sqrt{2}$ | $\dfrac{4m_s^2\sqrt{2}}{m\left(1-2v_g^2\right)}$ |
| $3$ | $4m_s$ <br><br> $\dfrac{8m_s}{\sqrt{1-4v_g^2}}$ | combishell, external | $2v_g$ | $n$ <br><br> $2m$ | $\dfrac{16m_s^2}{m\left(1-4v_g^2\right)}$ |

The second column gave the mass of the source and particle. The fourth gave the velocity of rotation and the fifth showed the relations between the quantum numbers of shell. Not all shells could be quantised simultaneously. Bohr's equations applied inside and outside each shell, and the physics could therefore be described using the Dirac and radiation equations in the space between two shells. The method of iteration, in which the wave accompanying the particle for one level became the position vector of the particle in the next, could be usefully described by a tiling or *tesseral hierarchy*,[4-6, 17] which we discuss further in section 8. We showed that the tesseral hierarchies applied in the quantum electromagnetic case,[10] as follows.

We may assemble a metric for a curved space that is electromagnetic in origin,

$$\mathrm{d}\tau^2 = \frac{2e_s}{r}\,\mathrm{d}s^2 - \frac{r}{2e_s}\,\mathrm{d}r^2 - r^2\left(\mathrm{d}\vartheta^2 + \sin^2\vartheta\,\mathrm{d}\varphi^2\right), \quad e_s = \frac{Ze^2}{m_e} \quad (6.\mathrm{A})$$

Bohr's equations, (3.A), for the interaction between a particle and this source, then follow from Bohr's equations, (5.C), and the second of equations (5.B) with $m_s$, $v_g$ and $h_g$ replaced by $e_s$, $v_e$ and $h$, respectively. We use the same procedure for quantisation as we used for gravity,[10] except that we suppose $m_e$ acquires a factor $1/\sqrt{1-v_e^2}$ on moving to the frame of the source, while $h$ is already defined in this frame. The Circular transformation could be derived by assuming that the circumference of circles centred on the origin suffered a contraction proportional to their radii.[8] If, in addition, the temporal co-ordinate suffers a similar dilation, we may suppose that both are caused by movement. The temporal, $s$, and a spatial, $r$, co-ordinate swapped each time we stepped up or down the table,[10] and we suppose that electromagnetism and gravity apply to an odd and an even shell, respectively. The appropriate Lorentz factor changes from $\sqrt{1-v_e^2}$ to $v_e$, which together form a four-vector. We may then use the electromagnetic equivalent of the second of equations (5.B).

For space 2 in the table, where the velocity was $v_g \sqrt{2}$, the Lorentz factor becomes $\sqrt{1-2v_g^2}$, which from the second of equations (5.B) gives the metric[10]

$$\mathrm{d}\tau^2 = \left(1 - 2m_s/r\right)\mathrm{d}s^2 - \frac{\mathrm{d}r^2}{\left(1 - 2m_s/r\right)} - r^2\left(\mathrm{d}\vartheta^2 + \sin^2\vartheta\,\mathrm{d}\varphi^2\right) \quad (6.B)$$

which is the metric predicted by General Relativity for the exterior of a spherical source of gravitation. We suppose that we are in space 2 for gravity. Other ways of performing this calculation are given by Bell and Diaz.[10] To return to General Relativity, we must imagine that the thin shell studied by Kuchar[22] and using canonical quantisation, by, for example, Berezin,[11] must be double, a single thin shell surrounded by another thin shell with the same centre. We reproduce their findings, in the limit as the two shells approach each other, excepting that for them the double shell expands or contracts, while here it spins. The temporal circumference of the shell and a radial co-ordinate may be interchanged, since this is another example of a transformation to a frame in which the particle becomes a tachyon.

## 7. General Relativity in the general case

Our account so far[7-10] has not included the full generality of Einstein's equation,[19] but only those instances where the energy-momentum-stress tensor may be represented by an energy-momentum four-vector. However, this did not limit the domain of the quantum theory to these instances.[10] When Einstein created General Relativity, he chose classical behaviour for the tensor on the grounds that insufficient was known about the actual interaction between particles, so he said in 1922 and afterwards.[19] In the case of quantum theory in the first twenty or so years of the 20th century he was clearly correct since quantum electrodynamics had not yet arrived. However, formulae giving statistical explanations for some classical laws, for example, the idea that pressure might be due to the impact of many particles, had already been found by Boltzmann and others during the 19th century.[15] Here electromagnetic forces are enlarged upon to explain, for example, the kinetic theory of gases. If we borrow this statistical theory, we may account for other more general instances of the tensor as being the collective behaviour of individual particles, each described by an energy-momentum four-vector. In this way the behaviour of electromagnetism and gravity may be made parallel. We have the same quantum theory for individual particles and the same generalisations for collections.

For example, the condition that the tensor is formed from a collection of four-vectors can be written[10]

$$T_{\mu\nu}(x_i) = \sum_{k=0}^{j} f_k(x_i) m_k v_{k\mu}(x_i) v_{k\nu}(x_i)$$

where the $T_{\mu\nu}$ are components of the energy-momentum-stress tensor, the $x_i$ label the points in the distribution, $m_k$ is the mass of the $k$th particle, $f_k m_k$ is the mass density at point $x_i$ in the rest frame, $v_{k\mu}$ is the component of a velocity four-vector and $j$, the number of particles, may be any positive integer. We require the equations to have a solution for some $f_k$, $m_k$ and $v_{k\mu}$. This equation is very underdetermined. For example, if we transform to the frame in which $T_{\mu\nu}$ is diagonal,[20] suppose our space two-dimensional and allow three particles, we obtain

$$\begin{pmatrix} T_{00} \\ T_{11} \\ 0 \end{pmatrix} = \begin{pmatrix} v_{00}^2 & v_{10}^2 & v_{20}^2 \\ v_{01}^2 & v_{11}^2 & v_{21}^2 \\ v_{00}v_{01} & v_{10}v_{11} & v_{20}v_{21} \end{pmatrix} \begin{pmatrix} m_0 f_0 \\ m_1 f_1 \\ m_2 f_2 \end{pmatrix}$$

If we regard $v_{k\mu}$ as given and the densities, $m_k f_k$, as unknown, we require the three by three matrix to have an inverse,[2] and may set the values of the $v_{k\mu}v_{k\nu}$ accordingly. If they are given real values, the $m_k f_k$ will have real values. $m_k$ must then be chosen so that the $f_k$ are positive. This might require $m_k$ to be negative, but this is permitted.[10] No new principles are involved in extending from a two-dimensional to a four-dimensional moving space and other approaches are also possible.

## 8.    History, physics and philosophy

Finally, since this is a conference on physical interpretations, we lay to rest the enmity between those ancient adversaries, Newton and Descartes. In book 2, propositions 52, 53 and the succeeding scholium,[24] Newton models fluid flow round a central spherical rotating body, given surroundings with spherical symmetry. Newton then discusses Descartes' vortex theory,[16] objecting that the vortex in such a fluid has a period proportional to the square of the distance from the centre of the sphere, while for gravity in the limit of increasingly rarefied air the period for a circular orbit is proportional to the distance to the power three halves. Further, he extends his argument to elliptical flow and objects that a fluid would flow faster at the aphelion of an ellipse than at the perihelion in circumstances which occur in our solar system. We have overcome the latter objection here by finding a sections through of spacetime in which the ellipse becomes a circle, while the conflict described in the former vanishes if we assume both of Bohr's equations, (5.C). The fluid obeys Bohr's second equation, while the rarefied air obeys Bohr's first, both in the frame of the source. We note that the satellites of larger bodies in the solar system do obey both equations approximately.[26]

Indeed, it is possible to derive both of equations (5.C) from a single proposition about a fluid on the assumption that the circular motion of a particle behaves in the way we described, splitting into two everywhere orthogonal circular components, with one circumference temporal and the other spatial.[8-10] We assume Newton's equation for the vortex in the frame of the particle,

$$2\pi rv = k$$

where $r$ is the radius, $v$ the velocity of the particle and fluid and $k$ is constant. We call this *the vortex equation*. We split the circumference into two circular components in two orthogonal planes in spacetime,

$$\frac{2\pi rv}{\sqrt{1-v^2}} = k_1, \quad \frac{2\pi rv^2}{\sqrt{1-v^2}} = k_2, \quad k_2^2 - k_1^2 = k^2 \qquad (8.\text{A})$$

The first equation describes another vortex, this time in the frame of the source, and we assume $k_1$ is constant, which leads to a constant $k_2$. If we set

$$k_1 = \frac{n_\theta h}{m_e} \text{ or } \frac{n_\theta h_g}{m_e}, \quad k_2 = \frac{Ze^2}{m_e} \text{ or } \frac{m_s}{\sqrt{1-v^2}}$$

we obtain Bohr's equations, (3.A) and (5.C). If we introduce curvature, the vortex described in the first of equations (8.A) can be specified using the co-ordinates for a single plane, leaving the second plane, described by the second equation, free for a spatial attraction governed by the geometrical properties of rays and the kinematical laws of circular motion[21, 24] in a flat two or three-dimensional space.[8, 9] We assumed a two-dimensional space, retrieving the third dimension from the radius of the vortex in the first equation,[7-10] but we could have insisted on a separate dimension instead, as in the de Sitter Universe, a five-dimensional sphere with spacetime as the surface.

Returning to the assumptions made throughout previously,[7-10] we conjecture that we may introduce a second index into the Dirac and radiation equations to describe the behaviour of $k_1$ and $k_2$ in a second copy of spacetime, each with four dimensions. We might hope for a link with the electro-weak theory in that case, since, if we associate $k_1$ with the temporal component and gravity, $k_2$ may have three components, inviting comparison. This implies the unification of $h$ and $h_g$ for each of the spaces between the shells in the original spacetime. If we wish to represent the scaling Bohr orbits as well, within a single theory, we may use tiles of different levels in a duplex tesseral hierarchy in four dimensions like the Hyper-cube or Hyper-HoR hierarchy.[3] The patterned tiles[23] may be addressed and changed using the arithmetic.[3, 8, 9] We conjecture that, instead, we might succeed in relating only electromagnetism and gravity using a single tesseral hierarchy by considering series of scaling Bohr orbits and scaling sets of series.

Newton also discusses the vortex of a fluid with cylindrical symmetry in, book 2, proposition 51.[24] Here he finds that the velocity of rotation is constant so that the period is proportional to the distance from the axis of symmetry. We have already determined in part 1 that in moving from the space formed by the Circular transformation back to the original, we should multiply distances measured along the rotating circumference by the length of the radius. We showed in Bell and Diaz[10] and in section 6 here that this could be seen as equivalent to removing a Lorentz contraction. This meant that we should also remove a Lorentz dilation from temporal components like the period in going from Newtonian mechanics to invariance under Special Relativity.[21, 24] This does not keep the velocity of rotation constant. If, however, the same factor could be applied to both the spatial and temporal components, it would. Work in progress shows that such a change can be induced if we use a Euclidean signature, $+ + + +$, rather than the Minkowski signature, $- + + +$, for spacetime. Appearances can still be saved, since the Dirac and radiation equations remain invariant. Then Newton's model of a cylindrical vortex would fit the cylindrical model used in part 1 and discussed in section 3 here.

## 9.    Summary

Roughly, we may summarise our results as follows:

| Paper | Assumption | Derived Result |
|---|---|---|
| 2000[7] | QED theory | A four-vector spinor |
| 2004pt1[8] | QED theory + 4-vector spinor | Bohr's equations |
| 2004pt2[8] | Bohr's equations | QED theory |
| 2004[9] | 2 orthogonal sets of Bohr's equations | QED theory |
| 2003[10] | General Relativity | Bohr's equations |
| 2004 here | Vortex equation | Bohr's equations |

**Acknowledgements**

One of us (Bell) would like to acknowledge the assistance of E.A.E. Bell. She would like to dedicate this paper to Grace Beryl Radley, who put up with so much without knowing why.